# Particle Filtering for Enhanced Parameter Estimation in Bilinear Systems Under Colored Noise

**Khalid Abd El Mageed Hag Elamin***

*Al Neelain University, College of Engineering, Khartoum, Sudan.*

**\*Corresponding Author**
Khalid, Abd El Mageed Hag Elamin, Al Neelain University, College of Engineering, Khartoum, Sudan.
Mail Id: kamelamin@neelain.edu.sd,  elaminkhalid32@gmail.com

**Abstract**
*This paper addresses the challenging problem of parameter estimation in bilinear systems under colored noise. A novel approach, termed B-PF-RLS, is proposed, combining a particle filter (PF) with a recursive least squares (RLS) estimator. The B-PF-RLS algorithm tackles the complexities arising from system nonlinearities and colored noise by effectively estimating unknown system states using the particle filter, which are then integrated into the RLS parameter estimation process. Furthermore, the paper introduces an enhanced particle filter that eliminates the need for explicit knowledge of the measurement noise variance, enhancing the method's practicality for real-world applications. Numerical examples demonstrate the B-PF-RLS algorithm's superior performance in accurately estimating both system parameters and states, even under uncertain noise conditions. This work offers a robust and effective solution for system identification in various engineering applications involving bilinear models subject to complex noise environments.*

**Keywords:** Bilinear Systems, Particle Filter, Parameter Estimation, Colored Noise, Optimal State Estimator

## 1. Introduction

In control systems, a bilinear system is a specific type of nonlinear system where the control input appears linearly but is multiplied by the state variables, creating a bilinear product term. Bilinear systems are particularly useful for describing processes where the effect of the control input on the system's state varies depending on the current state of the system [1]. For example, in a chemical reactor, the state of the reaction depends on both the concentration of reactants (state variables) and the temperature or pressure (control input) [2]. Similarly, the behavior of an electrical network with nonlinear components like diodes can be approximated by a bilinear system, where the state could be the voltage or current, and the control input could be an external voltage source [3].

The prominence of bilinear systems in numerous real-world control applications underscores the significance of the system identification process for these systems. This importance encourages control engineers to dedicate substantial attention to simulating and developing various models for such systems, aiming to achieve highly efficient and reliable system controllers [4]. A variety of methods for identifying bilinear systems have been introduced. For instance, a study on a filtering-based least-squares iterative algorithm has been conducted for the parameter estimation of bilinear systems affected by autoregressive noises [5]. Another study derived a state observer-based multi-innovation stochastic gradient algorithm and yet another introduced a bilinear state observer-based hierarchical least-squares method for bilinear state-space systems [6,7]. Identification of bilinear systems with colored noise using least-squares based iterative methods and maximum likelihood methods were developed by  and [ 8, 9].

One challenge in estimating bilinear system parameters is that the information vector in the parameter estimation algorithm may contain unknown system states. Deriving a state estimation algorithm for the bilinear system to replace the unknown state with its estimate can address this challenge [10]. Several studies have developed bilinear state observers using the Kalman filtering principle to effectively estimate unknown states and parameters. For example, propose a bilinear state observer-based recursive least squares algorithm for joint state-parameter estimation in bilinear systems, improving computational efficiency by decomposing the system into subsystems [11,12]. present an interactive estimation algorithm for unmeasurable states and parameters in bilinear systems with moving average noise, using a novel bilinear state observer and multi-innovation extended stochastic gradient algorithm [7].  proposed a bilinear state observer-based hierarchical multi-innovation stochastic gradient algorithm that effectively estimates parameters and states in bilinear systems, converging to their true values. developed optimal bilinear observers for bilinear state-space models, using interaction matrices to simplify the identification of models and observers from noisy measurements [13]. A stable bilinear observer can estimate the state of bilinear systems under any constant or nonconstant input, with the estimation error speed being independent of the applied input [14]. Minimal order state observers for bilinear systems can be designed without considering inputs, making the estimation error independent of inputs presented a design procedure for state observers in bilinear systems with bounded input, allowing for trade-offs between feedback amplification and input function bounds [15-17]. presented a new method for designing minimal order state-disturbance composite observers, which can effectively control bilinear systems, with applications in the headbox control system in the papermaking process.

The use of state observers in the model identification of bilinear systems presents several disadvantages, including the inapplicability of the Kalman filter, increased computational burden, challenges in handling measurement delays, and sensitivity to noise. These factors necessitate the development of specialized algorithms and models to ensure accurate and efficient state and parameter estimation [18,19]. Designing observers for discrete stochastic bilinear systems involves deriving mean square optimal linear unbiased observer equations, which can be sensitive to noisy output measurements, potentially affecting the accuracy of state reconstruction. This paper investigates the identification of bilinear system parameters influenced by different types of measurement noises such as white noise and colored noise based on a particle filtering approach. The particle filter is adapted with few particles, reducing the influence of distant observations on weight calculations, thereby reducing noise sensitivity [20].

**The main contributions of this paper are as follows:**
• The proposed algorithms in this paper achieve interactive state and parameter estimation for the bilinear system using the joint least squares principle combined with a particle filter state estimator.
• The particle filter is identified as the optimal state estimator for bilinear systems, effectively reducing noise sensitivity.
• The effects of colored measurement noises on the accuracy of bilinear system parameter estimation are investigated.
• In practical engineering applications, measuring output signals often entails dealing with output noise of unknown variance. Consequently, it becomes crucial to estimate the state, which is part of the information vector, under these circumstances. In such cases, the particle filter is optimized to calculate particle weights without requiring noise variance information. This modification makes the particle filter a more suitable choice compared to alternatives like the Kalman filter, which relies on known noise variance.

The layout of the remainder of this paper is as follows: Section 2 derives the identification model for the bilinear state-space models. In Section 3, we derive the particle filtering algorithm and present a particle weight calculation without knowing the measurement noise variance. A bilinear particle filter-based B-PF-RLS algorithm is developed to estimate the unknown parameters and states in Section 4. Numerical examples are shown in Section 5 to illustrate the benefits of the proposed methods in this paper. Finally, some concluding remarks are given in Section 6.

## 2. Identification Model for The Bilinear State-Space Models
Consider the following bilinear system in its observer canonical state-space form:

$$x(t+1) = Ax(t) + B\, x(t)u(t) + f\, u(t) + w(t), \tag{1}$$

$$y(t) = Hx(t) + e(t), \tag{2}$$

$$e(t) := \left(1 + k_1 q^{-1} + k_2 q^{-2} + \cdots + k_{n_k} q^{-n_k}\right)v(t). \tag{3}$$

Here, $v(t)$ is a white noise with zero mean, and $e(t)$ represents a colored noise [21].

Let's introduced some Notation and Assumptions for the Bilinear System

| SYMBOL | DESCRIPTION |
|---|---|
| $"A =: X"\ or\ "X := A"$ | Indicates that "$A$" is defined as "$X$" |
| $q$ | Unit back-shift operator, where $q^{-1}v(t)$ denotes $v(t-1)$ |
| $T$ | Superscript denoting the transpose of vectors/matrices |
| $w(t), v(t)$ | Process noise, measurement noise |
| $Q$ | Covariance matrix of $w(t)$, $Q \in \mathbb{R}^{n \times n}$ |
| $R$ | Covariance matrix of $v(t)$, $R \in \mathbb{R}$ |
| $p_0$ | is a large number $= 10^6 \gg 1$ |

**Assumptions:**
• The bilinear system described by equations (1) and (2) is stable, observable, and controllable.
• Noise processes w(t) and v(t) are uncorrelated and have the following properties:

$$E[w(t)] = 0, E[v(t)] = 0, E[w(t)v(i)] = 0$$

- The system is at rest for $t \leq 0$, i.e.: input: $u(t) = 0$, output: $y(t) = 0$, $x(t) = 0$, $w(t) = 0$ and $v(t) = 0$.

The matrices $A$, $B$, $f$ and $H$ representing system parameter are defined as follows:

$$A := \begin{bmatrix} -a_1 & 1 & 0 & \cdots & 0 \\ -a_2 & 0 & 1 & \cdots & 0 \\ \vdots & \vdots & \vdots & \ddots & \vdots \\ -a_{n-1} & 0 & 0 & \cdots & 1 \\ -a_n & 0 & 0 & \cdots & 0 \end{bmatrix} \in \mathbb{R}^{n \times n}, \; B := \begin{bmatrix} \boldsymbol{b}_1 \\ \boldsymbol{b}_2 \\ \vdots \\ \boldsymbol{b}_{n-1} \\ \boldsymbol{b}_n \end{bmatrix} \in \mathbb{R}^{n \times n} ; \boldsymbol{b}_1 \in \mathbb{R}^{1 \times n}, \; f := \begin{bmatrix} f_1 \\ f_2 \\ \vdots \\ f_{n-1} \\ f_n \end{bmatrix} \in \mathbb{R}^n$$

$$H := [1, \; 0, \; \cdots \; 0] \in \mathbb{R}^{1 \times n}. \tag{4}$$

The parameters $a_i$, $\boldsymbol{b}_i$, $j_i$ and $f_i$ are to be identified from the collected input $u(t)$ and output $y(t)$.

**Remark 1:** By transforming the bilinear state-space system presented in equations (1)-(2) into the observer canonical form, the identification process is simplified. This transformation effectively reduces the parameter space, resulting in more efficient and precise estimation.

Substituting these parameters into their matrix form within the system defined by equations (1) and (2), and applying straightforward transformations and manipulations detailed in reference the system can be rewritten as

$$x_1(t) = -\sum_{i=1}^n a_i \, x_1(t-i) + \sum_{i=1}^n \boldsymbol{b}_i \, x(t-i)u(t-i) + \sum_{i=1}^n f_i \, u(t-i) + \sum_{i=1}^n w_i(t-i)$$

$$y(t) = [1, \; 0, \; \cdots \; 0] \begin{bmatrix} x_1(t) \\ x_2(t) \\ \vdots \\ x_{n-1}(t) \\ x_n(t) \end{bmatrix} + e(t) \tag{5}$$

Using (3), system output can be written as

$$y(t) = x_1(t) + k_1 v(t-1) + k_2 v(t-2) + \cdots + k_{n_k} v(t-n_k) + v(t). \tag{6}$$

Substitute (5) in (6) yields

$$y(t) = -a_1 x_1(t-1), \ldots, -a_n x_1(t-n) + \boldsymbol{b}_1 x(t-1)u(t-1) \ldots + \boldsymbol{b}_n x(t-n)u(t-n) + f_1 u(t-1) + \cdots + f_n u(t-n) + J_1 v(t-1) + \cdots + J_n v(t-n) + \beta(t) \tag{7}$$

Where $\boldsymbol{b_n} = B(n, :)$.

Now, define:

$$\varphi_a(t) := [-x_1(t-1), \ldots, -x_1(t-n)]^T, \; \varphi_{xu}(t) := [x(t-1)^T u(t-1), \ldots, x(t-n)^T u(t-n)]^T,$$

$$\varphi_u(t) := [u(t-1), \ldots, u(t-n)]^T, \; \varphi_v(t) := [v(t-1), \ldots, v(t-n)]^T, \; \beta(t) = \sum_{i=1}^n w_i(t-i).$$

Based on equation (7) we define the information vector $\varphi(t)$ and the parameter vector $\theta$ as

$$\varphi(t) := [\varphi_a^T(t), \varphi_{xu}^T(t), \varphi_u^T(t), \varphi_v^T(t)]^T \text{ and } \theta = [a_1, \ldots, a_n, b_1, \ldots, b_n, f_1, \ldots, f_n, J_1, \ldots, J_n]^T \text{ respectively.}$$

Therefor, equation (7) can be rewritten as

$$y(t) = \varphi(t)^T \theta + \beta(t) + v(t) \tag{8}$$

Equation (8) is the identification model of the bilinear state-space system in (1) and (2).

**Remark 2:** Estimating the parameters of a bilinear system (defined by equations (1) and (2)) proves difficult due to the presence of unknown states $x(t-i)$ and noise $v(t-i)$ within the information vector $\varphi(t)$. This paper proposes a solution by integrating the recursive least squares identification technique with particle filtering. By leveraging this combined approach, we can effectively estimate both the system states and the parameter vector (containing $a_i$, $b_i$, $f_i$ and $k_i$) using only available input and output data.

## 3. Formulation of State Estimation Methods

This paper introduces a novel approach to bilinear system analysis, departing from the linearization techniques employed in previous methods, such as [23]. By utilizing a particle filter for state estimation, we directly handle the system's nonlinearities, eliminating the limitations of approximation. This framework also offers greater flexibility, allowing the incorporation of complex system models, even those with unknown parameters and measurement noise variances.

### 3.1 Particle Filter Algorithm

Particle filtering is a powerful technique for estimating the state of a dynamic system, especially when dealing with non-linear and non-Gaussian models. It employs a Monte Carlo approach, representing the probability distribution of the state with a set of weighted particles [24]. Here's a step-by-step breakdown of the algorithm:

**Step-1: Initialization:**
- **Number of particles:** Select the number of particles N.
- **Initial particles:** Sample initial particles $x_0^{(i)}$ from the prior distribution $p(x_0)$ for $i=1,2,\ldots,N$.
- **Initial weights:** Set the initial weights $w_0^{(i)} = \frac{1}{N}$ for all particles $i$.

**Step-2: Time update (Prediction step):**
- For each particle i at time t, propagate the state using the system dynamics:

$$x_{t+1}^{(i)} = Ax_t^{(i)} + Bx_t^{(i)}u_t + fu_t + w_t^{(i)} \tag{9}$$

where $w_t^{(i)}$ is a sample from the process noise distribution $p(w_t)$.

**Step-3: Measurement update (Correction step):**

- Compute the predicted measurement for each particle:
$$\hat{y}_t^{(i)} = cx_t^{(i)} \tag{10}$$
- Given $e(t) = J(q)v(t)$, calculate the measurement noise $e_t^{(i)}$:
$$e_t^{(i)} = y_t - \hat{y}_t^{(i)} \tag{11}$$

Here, $J(q)v(t)$ needs to be modeled. Assume $v(t)$ is white noise, then:
$$e_t^{(i)} = v_t + J_1 v_{t-1} + J_2 v_{t-2} + \cdots + J_n v_{t-n} \tag{12}$$
- Compute the likelihood of the measurement given the predicted state for each particle:
$$p(y_t|x_t^{(i)}) \propto \exp\left(-\frac{1}{2}(e_t^{(i)})^T \mathbf{R}^{-1} e_t^{(i)}\right) \tag{13}$$
where $R$ is the covariance matrix of the measurement noise $e(t)$.

**Step-4: Update weights:**

- Update the weights of each particle:
$$w_{t+1}^{(i)} = w_t^{(i)} p(y_t|x_t^{(i)}) \tag{14}$$

- Normalize the weights:
$$w_{t+1}^{(i)} = \frac{w_{t+1}^{(i)}}{\sum_{j=1}^{N} w_{t+1}^{(j)}} \tag{15}$$

**Step-5: Resampling:**

- Resample the particles to prevent degeneracy (where a few particles have almost all the weight):
  - Calculate the effective sample size:

$$N_{eff} = \frac{1}{\sum_{i=1}^{N} (w_{t+1}^{(i)})^2} \tag{16}$$

  - If $N_{eff}$ is less than a threshold $N_{threshold}$, resample the particles:
    - Perform resampling to generate a new set of particles $\{x_{t+1}^{(i)}\}$ by sampling with replacement from the current set of particles according to their weights.
    - Reset the weights of the resampled particles to $\frac{1}{N}$.

**Step-6: State Estimation:**
- Estimate the state at time $t$ as the weighted mean of the particles:

$$\hat{x}_{t+1} = \sum_{i=1}^{N} w_{t+1}^{(i)} x_{t+1}^{(i)} \tag{17}$$

**Step-7: Repeat:**
- Repeat steps 2 to 6 for each time step t.

**3.2 Particle Weight Calculation Without Knowing the Measurement Noise Variance**
The standard particle filter assumes known measurement noise variance, a limitation in real-world applications. This paper addresses this challenge by introducing a novel weight optimization method that directly estimates particle weights without relying on the explicit knowledge of noise variance.

**Problem:** Standard particle filters assume a known Gaussian distribution with a known variance for the measurement noise. However, in real-world scenarios, this variance is often unknown.

**Solution:** Based on the work in this paper proposes a modification that introduces a direct weight optimization method to address the issue of unknown measurement noise variance [25]. This method used Lagrange Multipliers for Constrained Optimization to directly estimates the weights of each particle, circumventing the need for explicit knowledge of the noise variance.
To solve this problem for the propose bilinear system we go through the following steps:

- **Particle Representation:**

A set of particles, $\hat{z}_j(t)$, are drawn from the predicted state distribution. The likelihood of observing the measurement $y(t)$ given a particle $\hat{x}_j(t)$ is denoted as $\Psi_j(t)$.

$$p\left(y(t) \middle| \hat{x}_j(t)\right) \coloneqq \Psi_j(t) \tag{18}$$

- **Approximation of Function:**

$$x(t) = A\, x(t-1) + B\, x(t-1) u(t-1) + f\, u(t-1) + e(t). \tag{19}$$

$$y(t) = g(x(t)) = c\, x(t) + k_1 v(t-1) + \cdots + k_{n_k} v(t - n_k) + v(t) \tag{20}$$

The goal is to approximate a non-linear function g(x(t)) using the particles and their associated weights. This approximation is given by:

$$g(x(t)) \approx \sum_{j=1}^{N} \Psi_j(t) g(\hat{x}_j(t)) \tag{21}$$

- **Weight Optimization:**

To find the optimal weights $\Psi_j(t)$, minimizes a cost function:

$$\left[\sum_{j=1}^{N} \Psi_j(t)(g(\hat{x}_j(t)) + e(t) - y(t))\right]^2 = \left(\sum_{j=1}^{N} \Psi_j(t) g(\hat{x}_j(t)) + \sum_{j=1}^{N} \Psi_j(t) e(t) - \sum_{j=1}^{N} \Psi_j(t) y(t)\right)^2 \tag{22}$$

This cost function represents the squared error between the predicted output based on the weighted particle values and the actual measurement.

- **Conditional Expectation:**

Taking the conditional expectation of the cost function (22) with respect to the measurement noise e(t) leads to:

$$E\left[\sum_{j=1}^{N} \Psi_j(t)(g(\hat{x}_j(t)) + e(t) - y(t))\right]^2 = \left(\sum_{j=1}^{N} \Psi_j(t) \gamma_j(t)\right)^2 + \sigma^2 \sum_{j=1}^{N} \Psi_j^2(t) \tag{23}$$

where $\gamma_j(t) = |y(t) - g(\hat{x}_j(t))|$ represents the absolute difference between the measurement and the predicted output for each particle.

- **Probability Minimization:**

Define $z(t)$ as a function of the weights and error terms. The goal is to find weights that minimize the probability of $z(t)$ being larger than a threshold $\gamma(t)$. This threshold is chosen as the maximum of all $\gamma_j(t)$ plus 1 [26].

$$\gamma(t) = \max\{\gamma_j(t), j = 1, \ldots, N\} + 1 \tag{24}$$

Assuming the measurement noise is Gaussian,
Given that $z(t)$ is a linear combination of Gaussian random variables (due to the Gaussian noise assumption), $z(t)$ itself follows a Gaussian distribution.
The mean $z_0$ of $z(t)$ is given by:

$$z_0 = \sum_{j=1}^{N} \Psi_j(t)\gamma_j(t) \tag{25}$$

The variance $\zeta^2$ of $z(t)$ is given by:

$$\zeta^2 = \sigma^2 \sum_{j=1}^{N} \Psi_j^2(t) \tag{26}$$

Since $z(t)$ is Gaussian with mean $z0$ and variance $\zeta^2$, the probability density function of $z(t)$ for $z \geq z_0$ is

$$p(z) = \begin{cases} \frac{2}{\sqrt{(2\pi)\zeta^2}} \exp\left(-\frac{(z-z_0)^2}{2\zeta^2}\right), z \geq z_0 \\ 0 \end{cases} \tag{27}$$

where $z_0 = \sum_{j=1}^{N} \Psi_j(t)\gamma_j(t)$ and $\zeta = \sigma\sqrt{\sum_{j=1}^{N} \Psi_j^2(t)}$.

Here, the factor of 2 accounts for the fact that we are considering the distribution only for $z \geq z_0$, effectively doubling the density in this region to maintain the correct normalization over the positive half. The density function (27) represents the likelihood of $z(t)$ given the weights $\Psi_j(t)$ and the Gaussian noise assumption. We are trying to optimize the weights $\Psi_j(t)$ in the context of a modified Particle Filter algorithm to minimize the probability that $z(t)$, a function of the weights and the errors, exceeds a threshold $\gamma(t)$.

**Key Definitions**

- $z(t) = \sum_{j=1}^{N} \Psi_j(t)\gamma_j(t)$
- $\gamma_j(t) = |y(t) - g(\hat{x}_j(t))|$
- $\gamma(t) = \max\{\gamma_j(t), j = 1, \ldots, N\} + 1$
- $\zeta = \sigma\sqrt{\sum_{j=1}^{N} \Psi_j^2(t)}$
- **Deriving the Optimization Problem**

We want to minimize the probability Prob ($z \geq \gamma(t)$). From the Gaussian density function (27). Consider the cumulative distribution function (CDF) of $z$:

$$Prob(z(t) \geq \gamma(t)) = \int_{\gamma(t)}^{\infty} p(z)dz \tag{28}$$

The smaller this probability, the better.
Minimizing (28) is equivalent to maximizing the argument inside the exponential of the Gaussian density function, as this will reduce the probability that $z(t)$ exceeds $\gamma(t)$. Therefore, we focus on maximizing the quantity:

$$\frac{\gamma(t) - z_0}{\zeta} = \frac{\gamma(t) - \sum_{j=1}^{N} \Psi_j(t)\gamma_j(t)}{\sigma\sqrt{\sum_{j=1}^{N} \Psi_j^2(t)}} \tag{29}$$

We need to find $\Psi_j(t)$ that maximize:

$$\frac{\gamma(t) - \sum_{j=1}^{N} \Psi_j(t)\gamma_j(t)}{\sqrt{\sum_{j=1}^{N} \Psi_j^2(t)}} \tag{30}$$

subject to the constraint $\sum_{j=1}^{N} \Psi_j(t) = 1$

To solve constrained optimization problem (30), we use the method of Lagrange Multipliers for Constrained Optimization. Define Lagrangian:

$$L(\Psi_j, \lambda) = \frac{\gamma(t) - \sum_{j=1}^{N} \Psi_j(t)\gamma_j(t)}{\sqrt{\sum_{j=1}^{N} \Psi_j^2(t)}} + \lambda\left(\sum_{j=1}^{N} \Psi_j(t) - 1\right) \tag{31}$$

here $\lambda$ is the Lagrange multiplier. Take the partial derivatives of $L$ with respect to $\Psi_j$ and set them to zero:

$$\frac{\partial L}{\partial \Psi_j} = \frac{-\gamma_j(t)}{\sqrt{\sum_{j=1}^{N}\Psi_j^2(t)}} - \frac{\left(\gamma(t)-\sum_{j=1}^{N}\Psi_j(t)\gamma_j(t)\right)\Psi_j}{\left(\sum_{j=1}^{N}\Psi_j^2(t)\right)^{\frac{3}{2}}} + \lambda = 0 \qquad (32)$$

Simplify and solve the resulting system (32) the optimal weights $\Psi_j(t)$ will be

$$\Psi_j(t) = \frac{\gamma(t)-\gamma_j(t)}{N\gamma(t)-\sum_{j=1}^{N}\gamma_j(t)} \qquad (33)$$

These weights are then used to update the particle weights in the Particle Filter algorithm.
Therefore, the weight is:

$$\omega_j(t) = \Psi_j(t)\omega_j(t-1) \qquad (34)$$

The weights $\omega_j(t)$ are then normalized to ensure that they sum to 1:

$$\bar{\omega}_j(t) = \frac{\omega_j(t)}{\sum_{k=1}^{N}\omega_k(t)} \qquad (35)$$

**Remark 3:** This modified particle filter uses a direct weight optimization approach to address the issue of unknown measurement noise variance. The key idea is to minimize a cost function that measures the error between the predicted and actual measurements, considering the uncertain noise variance. This method allows the filter to adapt to situations where the noise characteristics are unknown or change over time.

### 3.3 Bilinear State Observer for Bilinear Systems (BSO)

For comparative purposes to demonstrate the effectiveness of the proposed algorithm, this paper references common approaches to estimating the state of bilinear systems, such as in which typically rely on minimizing the state estimation error covariance matrix to obtain optimal states [10, 26, 27]. Inspired by this concept, a bilinear state observer for bilinear systems is derived using observation information. According to the state estimation part of a bilinear state observer algorithm (BSO-RLS) is implemented through the following iterative equations:

$$\hat{x}(t+1) = \hat{A}\hat{x}(t) + \hat{B}\hat{x}(t)u(t) + G\left[y(t) - H\hat{x}(t) - \hat{J}_1\hat{v}(t-1) - \cdots - \hat{J}_n\hat{v}(t-n)\right] \qquad (36)$$

$$G = [\hat{A}+\hat{B}u(t)]P(t)H^T[1+HP(t)H^T]^{-1} \qquad (37)$$

$$P(t+1) = [\hat{A}+\hat{B}u(t)]P(t)[\hat{A}+\hat{B}u(t)]^T - GHP(t)[\hat{A}+\hat{B}u(t)]^T \qquad (38)$$

Where, $P$ and $G$ represents estimation error covariance and observer gain respectively.
The algorithm (36) – (38) are combined with the parameter estimation techniques used in different studies to jointly estimate the state and parameters of a proposed bilinear system.

### 4. A Bilinear Particle Filter-Based B-PF-RLS Algorithm

This section presents an algorithm for jointly estimating parameters and states of a bilinear system with colored measurement noise. The algorithm combines a recursive least squares (RLS) estimator for parameter identification and a particle filter (PF) for state estimation with its weight calculated with a known measurement noise variance and unknown measurement noise variance. This approach effectively handles the challenges posed by the system's nonlinear nature and the presence of colored noise with and without knowing measurement noise variance.

### 4.1. The Parameter Estimation Algorithm
Define the quadratic criterion function as

$$C(\theta) := \sum_{j=1}^{L}\|y(j)-\varphi(j)^T\theta-\beta(j)\|^2, \qquad (39)$$

Based on the minimization of the criterion function (39), the system parameters are estimated according to the identification model (8) using least squares principle. Therefore, we have the following recursion relation [28].

$$\hat{\theta}(t) = \hat{\theta}(t-1) + L(t)\left[y(t)-\beta(t)-\varphi(t)^T\hat{\theta}(t-1)\right] \qquad (40)$$

$$L(t) = \frac{P(t-1)\varphi(t)}{1+\varphi(t)^T P(t-1)\varphi(t)} \tag{41}$$

$$P(t) = P(t-1) - L(t)[P(t-1)\varphi(t)]^T, P(0) = p_0 I_n, \tag{42}$$

There is a significant challenge in implementing algorithms (39)-(42) because the presence of $x(t-1)$, either fully or partially, in the formulations of $\varphi(t)$ defined in section 2 necessitates finding the actual or estimated elements of the vector $x(t-1)$. Similarly, the presence of the actual or estimated values of $w_i(t-i)$ and $v(t-i)$ complicates the formulations of both $\beta(t)$ and $\varphi_v(t)$, posing a significant challenge in formulating the information vector $\varphi(t)$, which is a crucial element in executing and implementing the mentioned algorithms. This issue can be overcome by incorporating the estimated state $\hat{x}(t-i)$ through the integration of the particle filtering algorithm within the overall execution of the algorithm to appropriately estimate the system state $x(t-1)$ and benefit from the advantages offered by the particle filter. Implementing and executing the particle filtering algorithm to estimate the system state $x(t-1)$ uses the system parameters to be estimated within the overall execution loop of the algorithm in an iterative manner, which is known as the idea of the auxiliary model mentioned in several previous studies such as and [8, 29]. As for $wi(t-i)$ and $v(t-i)$ mentioned in the formulations of $\beta(t)$ and $\varphi_v(t)$, the system parameters to be estimated can be used through the system equations under investigation according to the following equations

$$\hat{e}(t) = y(t) - H\hat{x}(t) \tag{43}$$

$$\hat{v}(t) = \hat{e}(t) - \hat{k}_1 \hat{v}(t-1) - \cdots - \hat{k}_{n_k} \hat{v}(t-n) \tag{43}$$

$$\hat{w}(t) = \hat{x}(t+1) - \hat{A}\hat{x}(t) - \hat{B}\hat{x}(t)u(t) - \hat{f}u(t) \tag{45}$$

Thus, the actual algorithm used becomes as follows

$$\hat{\theta}(t) = \hat{\theta}(t-1) + L(t)[y(t) - \hat{\beta}(t) - \hat{\varphi}(t)^T \hat{\theta}(t-1)] \tag{46}$$

$$L(t) = \frac{P(t-1)\hat{\varphi}(t)}{1+\hat{\varphi}(t)^T P(t-1)\hat{\varphi}(t)} \tag{47}$$

$$P(t) = P(t-1) - L(t)[P(t-1)\hat{\varphi}(t)]^T, P(0) = p_0 I_n, \tag{48}$$

Where $\hat{\varphi}(t) := [\hat{\varphi}_a^T(t), \hat{\varphi}_{xu}^T(t), \varphi_u^T(t), \hat{\varphi}_v^T(t)]^T$, $\hat{\beta}(t) = \sum_{i=1}^n \hat{w}_i(t-i)$.

## 4.2. The State Estimation Algorithm

The previous section dealt with the problem of the non-measurable states of the information vector $\varphi(t)$ and based on the idea of the auxiliary model replaced the unknown states and the unknown noises with their estimates. In this section we use the Particle filter algorithm in section 3.1 to estimate the system state. The following steps leads to obtain $\hat{x}(t-i)$.

**Step 1: Propagate particles using system model equation**

$$x_{particles}(t) = \hat{A} * x_{particles}(t-1) + \hat{B} * x_{particles}(t-1)u(t-1) + \hat{f}u(t-1) + \hat{w}(t-1) \tag{49}$$

**Step 2: Calculating weights**

$$z = H * x_{particles}(t) + \hat{e}(t) \tag{50}$$

where $z$ is the predicted measurements.

$$error = z - H * x_{particles}(t) + \hat{k}_1 \hat{v}(t-1) + \cdots + \hat{k}_{n_k}\hat{v}(t-n_k) \tag{51}$$

$$weights(number\ of\ particles) = \frac{1}{\sqrt{(2\pi)R^2}} \exp\left(-\frac{(error)^2}{2R}\right) \tag{52}$$

**Step 3: Normalize weights**

$$weights(:,t) = \frac{weights}{\sum weights} \tag{53}$$

**Step 4: Resampling**
As done in section 3.1 equation (16) to prevent degeneracy (where a few particles have almost all the weight).

**Step 5: Estimate system states**
Resampled indices = $f(weights, number\ of\ particles)$, i.e. its a function of weight and number of particles.
Resampled particles = $f(particles, resampled\ indices)$, i.e. its a function of particles and resampled indices.

Therefore,

$$\hat{x}(t+1) = \textbf{Mean}(\text{Resampled particles}) \tag{54}$$

**Remark 4:** A Particle filter estimates the state of the bilinear system under the assumption that the system parameters are known. To address this challenge, the idea of an auxiliary model that replaces all system parameters and system noises with their estimates in the particle propagate equation as shown in step 1 above.

**Remark 5:** In step 3 of the weight's calculation, the value of $R$ (measurement variance) must be known. However, in real-time practical applications, this value is typically unknown, posing a significant challenge for calculating the particles' weights. To address this issue, Section 3.2 introduces a direct weight optimization approach using Lagrange Multipliers for constrained optimization, which tackles the problem of unknown measurement noise variance.

## 4.3. The Joint Parameter and State Estimation Algorithm
Combining the parameter estimation algorithm in (43)-(48) with the state estimation algorithm in (49)-(54), we obtain a bilinear particle filter based recursive-least squares algorithm (B-PF-RLS) to combined estimated state with the estimated bilinear system parameter vectors.

$$\hat{\theta}(t) = \hat{\theta}(t-1) + L(t)\left[y(t) - \hat{\beta}(t) - \hat{\varphi}(t)^T \hat{\theta}(t-1)\right] \tag{55}$$

$$L(t) = \frac{P(t-1)\hat{\varphi}(t)}{1+\hat{\varphi}(t)^T P(t-1)\hat{\varphi}(t)} \tag{56}$$

$$P(t) = P(t-1) - L(t)[P(t-1)\hat{\varphi}(t)]^T, P(0) = p_0 I_n, \tag{57}$$

$$\hat{\varphi}_a(t) := [-\hat{x}_1(t-1), \ldots, -\hat{x}_1(t-n)]^T \tag{58}$$

$$\hat{\varphi}_{xu}(t) := [\hat{x}(t-1)^T u(t-1), \ldots, \hat{x}(t-n)^T u(t-n)]^T \tag{59}$$

$$\varphi_u(t) := [u(t-1), \ldots, u(t-n)]^T \tag{60}$$

$$\hat{\varphi}_v(t) := [\hat{v}(t-1), \ldots, v(t-n)]^T \tag{61}$$

$$\hat{\beta}(t) = \sum_{i=1}^{n} \hat{w}_i(t-i) \tag{62}$$

$$\hat{\varphi}(t) := [\hat{\varphi}_a^T(t), \hat{\varphi}_{xu}^T(t), \varphi_u^T(t), \hat{\varphi}_v^T(t)]^T \tag{63}$$

$$\hat{\theta} = \left[\hat{a}_1, \ldots, \hat{a}_n, \hat{b}_1, \ldots, \hat{b}_n, \hat{f}_1, \ldots, \hat{f}_n, \hat{k}_1, \ldots, \hat{k}_{n_k}\right]^T \tag{64}$$

$$particles(:,:) = \hat{A}\ particles(:,:,t) + \hat{B}\ particles(:,:,t)u(t) + \hat{f}\ u(t) + \hat{w}(t) \tag{65}$$

$$error = z - c\ particles(:, number\ of\ particles, t) + \hat{k}_1 \hat{v}(t-1) + \cdots + \hat{k}_{n_k} \hat{v}(t-n_k) \tag{66}$$

• **If measurement noise variance known**

$$weights = \frac{1}{\sqrt{(2\pi)R^2}} \exp\left(-\frac{(error)^2}{2R}\right) \tag{67}$$

Resampled indices = $F(weights, number\ of\ particles)$ i.e. its a function of weight and number of particles.
Resampled particles = $F(particles, resampled\ indices)$ i.e. its a function of particles and resampled indices.

$$\hat{x}(t+1) = \textbf{Mean}(\text{Resampled particles}) \tag{68}$$

- **If measurement noise variance unknown**

$$\Psi_j(t) = \frac{\gamma(t)-\gamma_j(t)}{N\,\gamma(t)-\sum_{j=1}^{N}\gamma_j(t)}, \qquad \text{weights } = \Psi_j(t). \tag{69}$$

Resampled indices = $F(weights, number\ of\ particles)$ i.e. its a **Function** of weight and number of particles.
Resampled particles = $F(particles, resampled\ indices)$ i.e. its a **Function** of particles and resampled indices.

$$\hat{x}(t+1) = \mathbf{Mean}(\text{Resampled particles}) \tag{70}$$

$$\hat{e}(t) = y(t) - H\,\hat{x}(t) \tag{71}$$

$$\hat{v}(t) = \hat{e}(t) - \hat{k}_1 \hat{v}(t-1) - \cdots - \hat{k}_{n_k}\hat{v}(t-n_k) \tag{72}$$

$$\widehat{w}(t) = \hat{x}(t+1) - \hat{A}\hat{x}(t) - \hat{B}\hat{x}(t)u(t) - \hat{f}u(t) \tag{73}$$

$$\hat{A} := \begin{bmatrix} -\hat{a}_1(t) & 1 & 0 & \cdots & 0 \\ -\hat{a}_2(t) & 0 & 1 & \cdots & 0 \\ \vdots & \vdots & \vdots & \ddots & \vdots \\ -\hat{a}_{n-1}(t) & 0 & 0 & \cdots & 1 \\ -\hat{a}_n(t) & 0 & 0 & \cdots & 0 \end{bmatrix}, \hat{B}(t) = \begin{bmatrix} \hat{b}_1(t) \\ \hat{b}_2(t) \\ \hat{b}_3(t) \\ \vdots \\ \hat{b}_n(t) \end{bmatrix}, \hat{f}(t) = \begin{bmatrix} \hat{f}_1(t) \\ \hat{f}_2(t) \\ \hat{f}_3(t) \\ \vdots \\ \hat{f}_n(t) \end{bmatrix}. \tag{74}$$

**Remark 6:** State estimation algorithm use equation (68) with different values of measurement noise variance or use (70) to deal with unknown measurement noise variance to obtain state estimates exploited in the parameter estimation process. The parameter estimates are improved by using a specified value of the number of particles in the state estimation process to obtain a minimized state estimation error to improve the accuracy of the parameter estimation.

## 5. Numerical Examples
**Example 1:** Consider the following bilinear state - space system in its observable-canonical form

$$x(t+1) = Ax(t) + G\,x(t)u(t) + F\,u(t) + w(t),$$
$$y(t) = Hx(t) + e(t),$$
$$e(t) = k_1 v(t-1) + k_2 v(t-2) + v(t).$$

$$A = \begin{bmatrix} -a_1 & 1 \\ -a_2 & 0 \end{bmatrix} = \begin{bmatrix} -0.30 & 1 \\ 0.25 & 0 \end{bmatrix}$$
$$G = \begin{bmatrix} g_{11} & g_{12} \\ g_{21} & g_{22} \end{bmatrix} = \begin{bmatrix} 0.10 & 0.15 \\ 0.30 & 0.20 \end{bmatrix}, \quad H = [1,\ 0],$$
$$F = \begin{bmatrix} f_1 \\ f_2 \end{bmatrix} = \begin{bmatrix} 1.15 \\ 1.56 \end{bmatrix}, \quad w(t) = \begin{bmatrix} w_1(t) \\ w_2(t) \end{bmatrix},$$

The parameter vector to be identified is given by:

$$\theta = [a_1, a_2, g_{11}, g_{12}, g_{21}, g_{22}, f_1, f_2, k_1, k_2]^T,$$

$$= [0.30, -0.25, 0.10, 0.15, 0.30, 0.20, 1.15, 1.56, -0.14, 0.20]^T.$$

- When modelling, system parameters should ensure the stability, controllability, and observability of the system. In the simulation, the input $\{u(t)\}$ is a pseudo-random binary sequence generated by the Matlab function $u$ = idinput ([8191,1,1], prbs' ',[0,1], [-1,1]), $w_1(t)$ and $w_2(t)$ are random noise sequences with zero mean and variance $\sigma^2_{w_1} = 0.07^2$, and $\sigma^2_{w_2} = 0.01^2$ respectively. $v(t)$ is a random noise sequence with zero mean and variance $\sigma^2_v = 0.45^2$, $\sigma^2_v = 0.50^2$, $\sigma^2_v = 0.80^2$ and $\sigma^2_v = 1.00^2$. The data length $L$= 3000 is set, and different values are chosen for the number of particles and the measurement noise $\sigma^2_v$. System parameter and state estimates are generated by applying the B-PF-RLS algorithm.
- The parameters estimates and errors $\delta_\theta = \|\hat{\theta} - \theta\|/\|\theta\|$ at $\sigma^2_v = 0.45^2$, $0.8^2$, and $1.0^2$ are summarized in Table 1. According to these different values of the measurement noise variance the parameter estimation errors are plotted against $t$ in Fig. 1. Figure (2) and Table 2 shows the $\sigma^2_v = 0.8^2$ parameter estimation error of the proposed algorithm compared to the BSO-RLS algorithm. The noise estimates $\hat{v}(t), \widehat{w_1}(t)$ and $\widehat{w_2}(t)$ for B-PF-RLS with $\sigma^2_v = 0.8^2$ are shown in figure (3).
- The true state $x_1$ and $x_2$ with their estimates $\hat{x}_1$ and $\hat{x}_2$ and their associated errors for $\sigma^2_v = 0.8^2$ and 1002 particles are shown in Fig. (4). The collected input and output data are shown in Figure (5).

- The distribution of particles in the state space, with $x_1$ on the x-axis and $x_2$ on the y-axis with the weights of each particle at time t are shown in figure (6). Figure (7) illustrates the probability density of the particles' values at different time steps. Furthermore, the figure displays the weights of the particles over time using a heatmap.
- The probability density functions (PDFs) of measurement values of the observed measurement is shown in figure (8).
- Figure (9) compares the root mean square error for state $x_1$ for B-PF-RLS algorithm with 1002 particles and 2000 particles.
- The Root mean -square error for state $x_1$ with particle weight calculated with known $\sigma_v^2$ and unknown $\sigma_v^2$ is illustrated in figure (10).

Looking at Tables 1-2 and Figures 1-10, we can draw some conclusions from these tables and figures.
- As can be seen from Table 1, and figure (1) the parameter estimation error $\delta_\theta$ increase as the measurement noise variance $\sigma_v^2$ increases and vice versa.
- The best estimate of the state is obtained when the root mean -square error $\delta_x$ between the true state and estimated state decreases with the increase of number of particles. This is reflected in the improved accuracy of the parameter estimates presented by the B-PF-RLS algorithm. See figures (9).
- We find that the B-PF-RLS algorithm provides better parameter estimation accuracy than the BSO-RLS algorithm under the same conditions, which makes this algorithm efficient and robust. See Figure (2) and Table 2.
- The close clustering of particles indicates that the filter has converged to a highly certain state estimate. This tight grouping suggests that the particles are accurately tracking the system's dynamics, leading to a high level of confidence in the state estimation. The distribution confirms that the filter is accurately capturing the true state of the system, highlighting the effectiveness and reliability of the particle filter see figure (6).
- Most of the particles having higher weights as in figure (6) indicating they are close to the true state. This distribution suggests the filter is performing well, with high-weight particles contributing most to the state estimate. The resampling step will maintain diversity by replicating high-weight particles and removing low-weight ones, improving accuracy and reliability of state estimates. Particle filter is confirmed to offer precise and reliable state estimates.
- Figure (7) show that the particles are consistently concentrated around the true state values. This indicates that the filter accurately captures and maintains the true state over time, providing reliable and precise state estimates. The heatmap of particle weights demonstrates distinct bands of higher weights, indicating that specific particles consistently have higher weights and contribute significantly to the state estimate. This confirms that the filter effectively identifies and tracks the true state by prioritizing the most relevant particles, maintaining diversity and accuracy.
- For the actual output $y$ and the observed measurement value $z$ depicted in figure (8), the alignment of the peaks with expected values indicates that the particle filter accurately estimated state and measurement values, resulting in highly accurate PDFs. This demonstrates the filter's effectiveness in capturing true measurement distributions and confirming the reliability of state estimates. The figure serves as strong evidence of the filter's ability to provide precise and reliable measurement estimates, validating its performance in accurately tracking system dynamics.

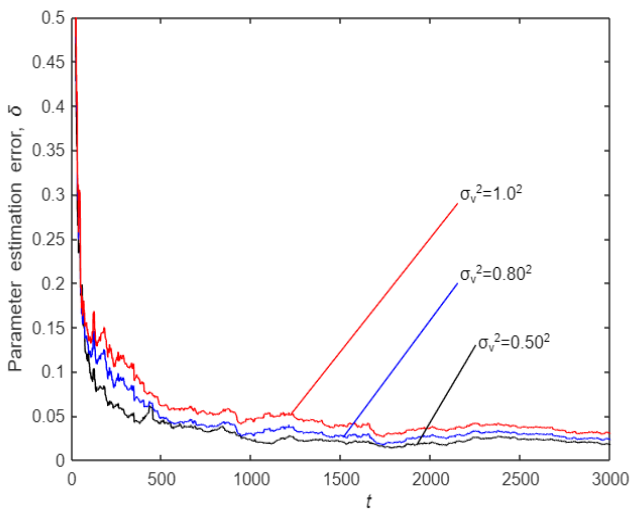

**Figuere 1:** The B-PF-RLS estimation errors $\delta_\theta$ for $R_v$ =0.50, 0.80 and 1.00, 1002 Particles, $Q_w = [0.20, 0.01]I_2$).

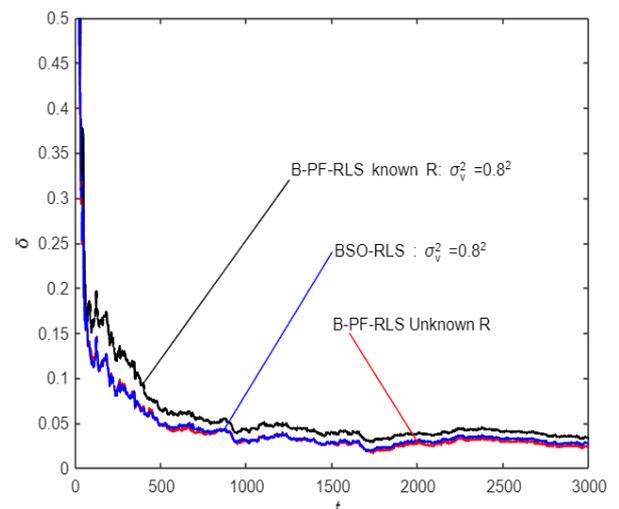

**Figure 2:** The B-PF-RLS estimation error $\delta_\theta$ against t compared to BSO-RLS algorithm for 1002 number of particles, $R_v = 0.8$, $Q_w = [0.07, 0.01]I_2$).

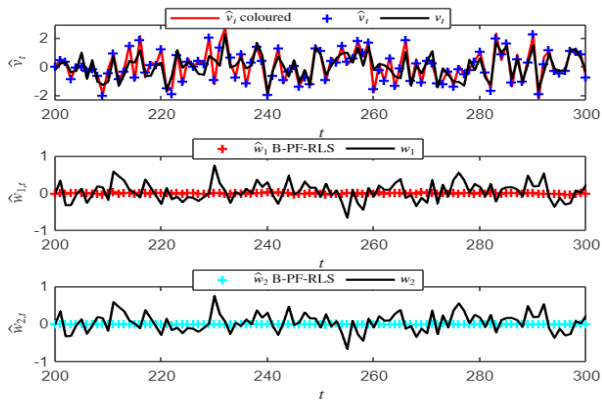

**Figure 3:** The estimates of $w_1(t)$, $w_2(t)$, and $v(t)$ for B-PF-RLS for $R_v = 0.80$, $Q_w = [0.07, 0.01]I_2$).

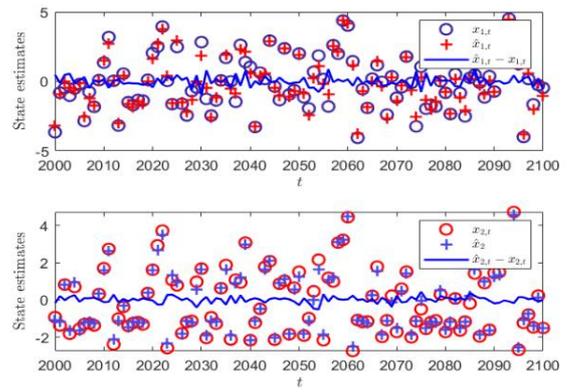

**Figure 4:** The B-PF-RLS state estimates of $x_{1t}$ and $x_{2t}$ with the deviation $\hat{x}_{1t} - x_{1t}$ and $\hat{x}_{2t} - x_{2t}$ against $t$ ($R_v = 0.8$, 1002 particles).

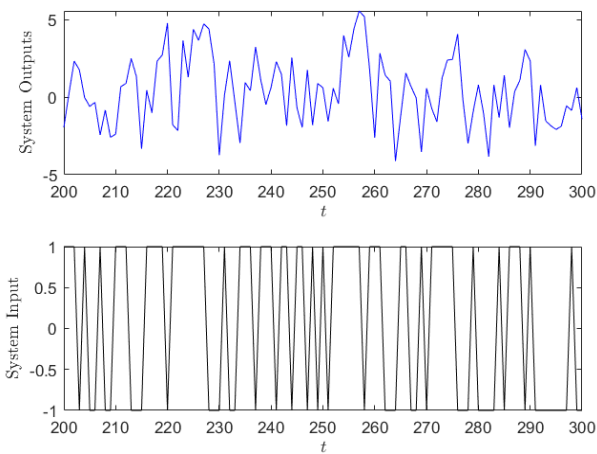

**Figure 5:** The input $u(t)$ and output $y(t)$ collected data used in example 1

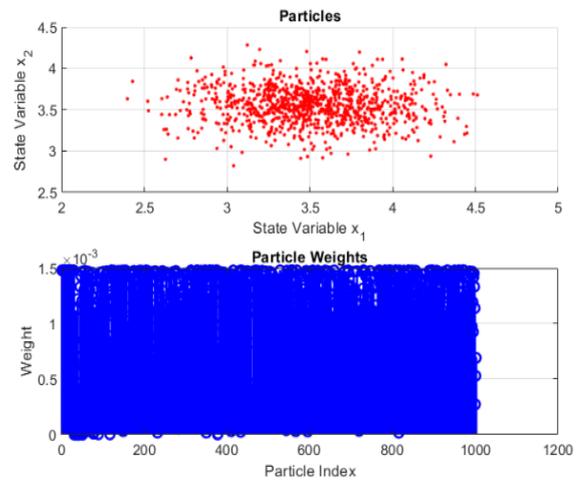

**Figure 6:** The distribution of particles in the state space, with the weights of each particle at time t

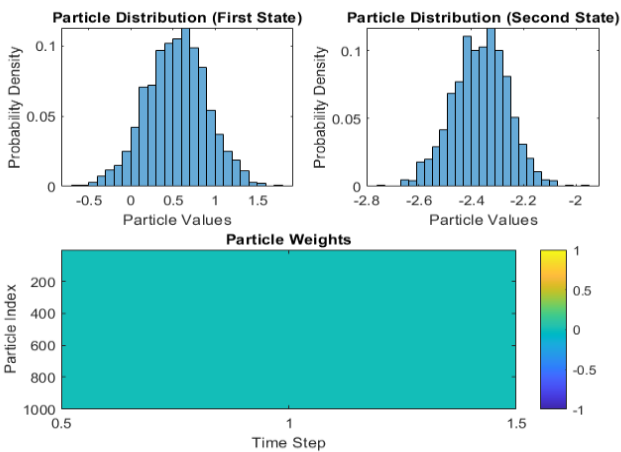

**Figure 7:** The probability density of the particles' values at different time step with their weights using a heatmap

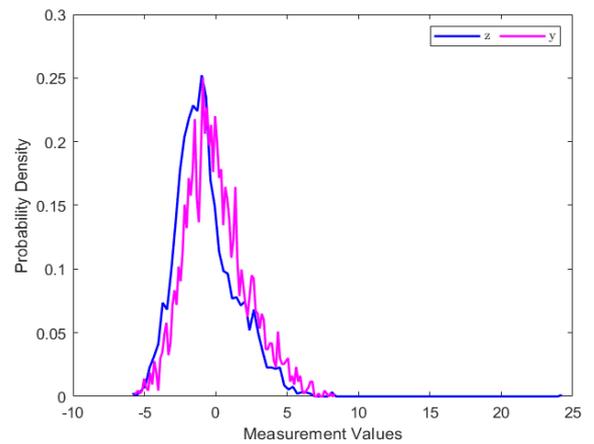

**Figure 8:** The Measurement Probability Density Functions (PDFs) of measurement values of the observed measurement

| $\sigma_v^2$ | t | $a_1$ | $a_2$ | $b_{11}$ | $b_{12}$ | $b_{21}$ | $b_{22}$ | $f_1$ | $f_2$ | $k_1$ | $k_2$ | $\delta_\theta\%$ |
|---|---|---|---|---|---|---|---|---|---|---|---|---|
| $0.45^2$ | 100 | 0.2554 | -0.2525 | 0.1827 | -0.0493 | 0.2795 | 0.0278 | 1.1034 | 1.6469 | -0.0417 | -0.0120 | 15.2134 |
|  | 1000 | 0.3009 | -0.2456 | 0.1169 | 0.1106 | 0.2978 | 0.1789 | 1.1316 | 1.5678 | -0.0952 | -0.0218 | 3.5082 |
|  | 3000 | 0.2904 | -0.2508 | 0.1098 | 0.1320 | 0.2979 | 0.1948 | 1.1603 | 1.5341 | -0.1394 | 0.0268 | 1.8143 |
| $0.80^2$ | 100 | 0.1623 | -0.1843 | 0.0932 | 0.1546 | 0.1178 | 0.0546 | 1.1455 | 1.4553 | -0.1765 | 0.1095 | 15.6477 |
|  | 1000 | 0.2692 | -0.2287 | 0.1135 | 0.1349 | 0.2725 | 0.1885 | 1.1339 | 1.5191 | -0.1937 | 0.0188 | 4.2504 |
|  | 3000 | 0.2791 | -0.2490 | 0.1076 | 0.1414 | 0.2948 | 0.1962 | 1.1632 | 1.5130 | -0.1764 | 0.0292 | 3.3584 |
| $1.00^2$ | 100 | 0.1482 | -0.1732 | 0.0613 | 0.2087 | 0.0749 | 0.0476 | 1.1425 | 1.4431 | -0.1943 | 0.1204 | 18.3677 |
|  | 1000 | 0.2396 | -0.2148 | 0.1137 | 0.1465 | 0.2500 | 0.1915 | 1.1349 | 1.4788 | -0.2126 | 0.0283 | 7.0160 |
|  | 3000 | 0.2689 | -0.2474 | 0.1077 | 0.1465 | 0.2897 | 0.1970 | 1.1646 | 1.4971 | -0.1875 | 0.0291 | 4.4183 |
| True value |  | 0.3 | -0.25 | 0.1 | 0.14 | 0.3 | 0.2 | 1.15 | 1.56 | -0.14 | 0.01 |  |

**Table 1** The parameter estimates and errors of the B-PF-RLS algorithm for $\sigma_v^2 = 0.45^2$, $\sigma_v^2 = 0.80^2$ and $\sigma_v^2 = 1.00^2$, 1002 Particles

| Algo. | t | $a_1$ | $a_2$ | $b_{11}$ | $b_{12}$ | $b_{21}$ | $b_{22}$ | $f_1$ | $f_2$ | $k_1$ | $k_2$ | $\delta_\theta\%$ |
|---|---|---|---|---|---|---|---|---|---|---|---|---|
| B-PF RLS | 100 | 0.2197 | -0.2122 | 0.1304 | 0.0778 | 0.1297 | 0.0903 | 1.1261 | 1.5620 | -0.1488 | 0.0711 | 11.9220 |
|  | 1000 | 0.2889 | -0.2360 | 0.1278 | 0.1145 | 0.2709 | 0.1931 | 1.1346 | 1.5493 | -0.1541 | -0.0101 | 2.9635 |
|  | 3000 | 0.2872 | -0.2503 | 0.1146 | 0.1330 | 0.2929 | 0.1970 | 1.1662 | 1.5248 | -0.1571 | 0.0163 | 2.3819 |
| BSO-RLS | 100 | 0.2207 | -0.2176 | 0.0935 | 0.1052 | 0.1464 | 0.0666 | 1.1241 | 1.5630 | -0.1682 | 0.0584 | 11.4723 |
|  | 1000 | 0.2827 | -0.2398 | 0.1108 | 0.1250 | 0.2746 | 0.1896 | 1.1289 | 1.5406 | -0.1647 | -0.0195 | 3.0479 |
|  | 3000 | 0.2812 | -0.2520 | 0.1078 | 0.1368 | 0.2923 | 0.1974 | 1.1624 | 1.5166 | -0.1636 | 0.0104 | 2.7494 |
| True value |  | 0.3 | -0.25 | 0.1 | 0.14 | 0.3 | 0.2 | 1.15 | 1.56 | -0.14 | 0.01 |  |

**Table 2** The parameter estimates and errors of the B-PF-RLS algorithm with unknown $R_v$ compared to BSO-RLS algorithm, $R_v = 0.8$, $Q_w = [0.07, 0.01]I_2$).

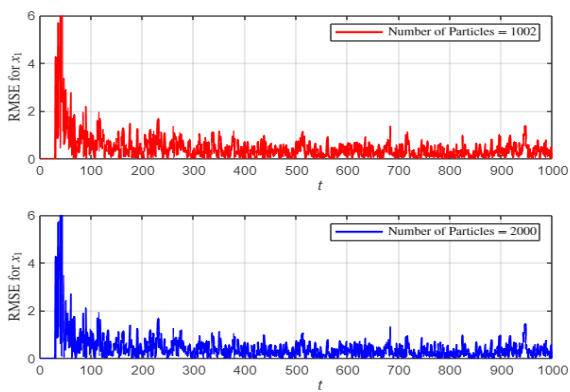

**Figure 9:** The root mean square error for state $x_1$ for 1002 particles and 2000 particles, ($\sigma_v^2 = 0.80^2$ and $Q_w = [0.07, 0.01]I_2$)

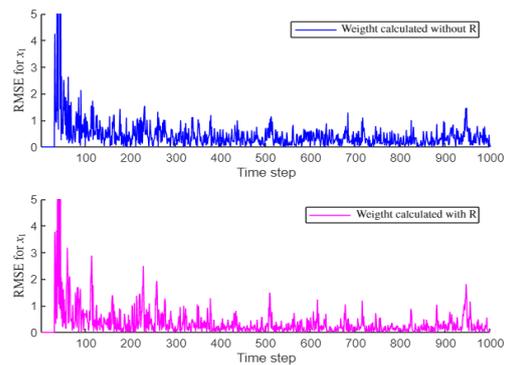

**Figure 10:** The root mean -square error for state $x_1$ with particle weight calculated with known $\sigma_v^2$ and unknown $\sigma_v^2$.

**Example 2**
To evaluate the effectiveness of the proposed algorithm on a large-scale system, a fourth-order bilinear state-space system is considered with colored noise for practical considerations, as follows:

$$x_{t+1} = \begin{bmatrix} -0.40 & 1 & 0 & 0 \\ 0.24 & 0 & 1 & 0 \\ 0.16 & 0 & 0 & 1 \\ -0.05 & 0 & 0 & 0 \end{bmatrix} x_t + \begin{bmatrix} -0.45 & 0.32 & 0.18 & -0.10 \\ -0.02 & 0.10 & -0.07 & 0 \\ 0.10 & -0.05 & 0 & 0.20 \\ 0.05 & 0 & 0.30 & -0.20 \end{bmatrix} x_t u_t + \begin{bmatrix} 1.20 \\ 1.60 \\ 0.60 \\ 2.12 \end{bmatrix} u_t + w_t$$

$$y_t = [1,0,0,0]x_t - 0.41 v(t-1) + v(t)$$

The parameter vector to be identified is
$$\theta = [a_1, a_2, a_3, a_4, b_{11}, b_{12}, b_{12}, b_{12}, b_{12}, b_{12}, b_{12}, b_{12}, b_{12}, b_{12}, b_{12}, b_{12}, b_{12}, b_{12}, b_{12}, f_1, f_2, f_2, f_2, k_1]^T$$

$$= \begin{bmatrix} 0.40, -0.24, -0.16, 0.05, -0.45, 0.32, 0.18, -0.10, -0.02, 0.10, -0.07, 0, 0.10, -0.05, 0, 0.20, 0.05, 0, 030, \\ -0.20, 1.20, 1.60, 0.60, 2.12, -0.41 \end{bmatrix}^T$$

In the simulation, the input $\{u(t)\}$ is a pseudo-random binary sequence generated by the MATLAB function. The process noise sequences $w_1(t), w_2(t), w_3(t)$ and $w_4(t)$ are random with zero mean and variances $\sigma_{w_1}^2 = 0.07^2$, $\sigma_{w_2}^2 = 0.01^2$, $\sigma_{w_3}^2 = 0.02^2$ and $\sigma_{w_4}^2 = 0.04^2$ respectively. The noise $v(t)$ is a white random sequence with zero mean and variances $\sigma_v^2 = 0.30^2$, $\sigma_v^2 = 0.80^2$ and $\sigma_v^2 = 1.0^2$. The data length $L$ is set to 5000. Under the measurement noise variances $\sigma_v^2 = 0.30^2$, $\sigma_v^2 = 0.80^2$ and $\sigma_v^2 = 1.0^2$, the B-PF-RLS algorithm is applied to estimate the parameters and states of this fourth-order bilinear system. There are 25 parameters to be identified, with their true values and estimates presented in Table 3. The error curves under different variances are shown in Figure 11. For the noise variance $\sigma_v^2 = 0.80^2$, parameter estimates over time are displayed in Figure 12, showing initial fluctuations but convergence to true values as $t$ increases. To illustrate the probabilistic characteristics of the method, the parameter estimates, Mean Absolute Deviation (MAD), and Root Mean Squared Deviation (RMSD) of the B-PF-RLS algorithm are summarized in Table 3. It is evident that the average parameter estimates closely match their true values. From Table 3-4 and Figures 11-12, it is observed that parameter estimates are close to their true values and the estimation accuracy improves as noise variances decrease.

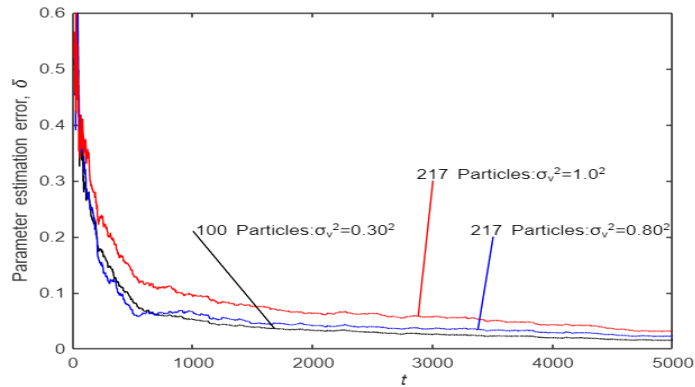

**Figure 11:** The B-PF-RLS estimation errors – against $t$ with $\sigma_v^2 = 0.30^2$, $\sigma_v^2 = 0.80^2$ and $\sigma_v^2 = 1.0^2$

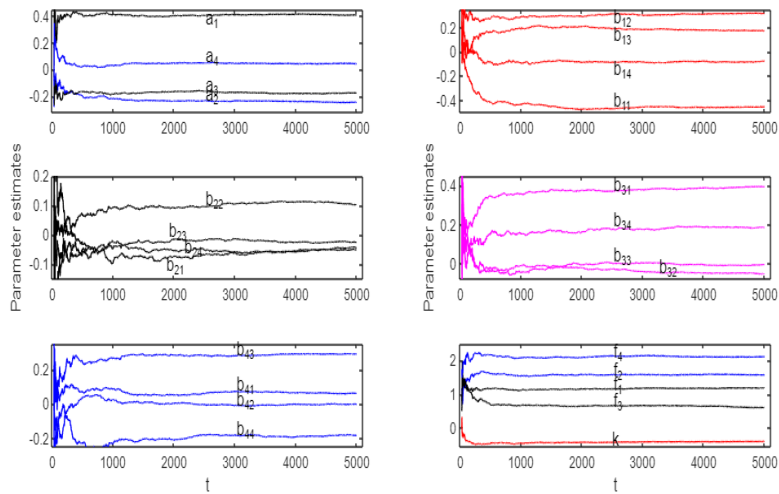

**Figure 12:** The B-PF-RLS parameters estimates against $t$ ( $\sigma_v^2 = 0.80^2$, 217 *particles*)

| Parameter | $\hat{\theta}$ | Mean | MAD | RMSD | True values |
|---|---|---|---|---|---|
| $a_1$ | 0.4087 | 0.40 | ±0.0168 | ±0.0587 | 0.40 |
| $a_2$ | -0.2432 | -0.22 | ±0.0211 | ±0.0370 | -0.24 |
| $a_3$ | -0.1752 | -0.17 | ±0.0099 | ±0.0281 | -0.16 |
| $a_4$ | 0.0444 | 0.05 | ±0.0093 | ±0.0330 | 0.05 |
| $b_{11}$ | -0.4527 | -0.43 | ±0.0420 | ±0.0834 | -0.45 |
| $b_{12}$ | 0.3139 | 0.30 | ±0.0169 | ±0.0369 | 0.32 |
| $b_{13}$ | 0.1716 | 0.18 | ±0.0227 | ±0.0483 | 0.18 |
| $b_{14}$ | -0.0779 | -0.08 | ±0.0159 | ±0.0368 | -0.10 |
| $b_{21}$ | -0.0251 | -0.03 | ±0.0121 | ±0.0310 | -0.02 |
| $b_{22}$ | 0.1026 | 0.09 | ±0.0192 | ±0.0378 | 0.10 |
| $b_{23}$ | -0.0497 | -0.04 | ±0.0219 | ±0.0475 | -0.07 |
| $b_{24}$ | -0.0457 | -0.06 | ±0.0145 | ±0.0276 | 0.00 |
| $b_{31}$ | 0.3945 | 0.36 | ±0.0331 | ±0.0640 | 0.40 |
| $b_{32}$ | -0.0542 | -0.03 | ±0.0214 | ±0.0549 | -0.05 |
| $b_{33}$ | -0.0032 | -0.01 | ±0.0155 | ±0.0299 | 0.00 |
| $b_{34}$ | 0.1871 | 0.17 | ±0.0218 | ±0.0416 | 0.20 |
| $b_{41}$ | 0.0604 | 0.07 | ±0.0112 | ±0.0216 | 0.05 |
| $b_{42}$ | -0.0013 | -0.00 | ±0.0156 | ±0.0417 | 0.00 |
| $b_{43}$ | 0.2903 | 0.27 | ±0.0233 | ±0.0474 | 0.30 |
| $b_{44}$ | -0.1817 | -0.19 | ±0.0241 | ±0.0480 | -0.20 |
| $f_1$ | 1.1890 | 0.67 | ±0.0553 | ±0.1434 | 1.20 |
| $f_2$ | 1.5758 | 1.55 | ±0.0533 | ±0.1776 | 1.60 |
| $f_3$ | 0.6074 | 0.67 | ±0.0553 | ±0.1434 | 0.60 |
| $f_4$ | 2.1112 | 2.08 | ±0.0615 | ±0.2311 | 2.12 |
| $k$ | -0.3987 | -0.41 | ±0.0258 | ±0.0643 | -0.41 |

**Table 3:** B-PF-RLS parameter estimates Mean, Absolute Deviation (MAD) and Root Mean Squared Deviation (RMSD) for 5000 runs

| $\sigma_v^2$ | $a_1 = 0.40$ | $a_2 = -0.24$ | $a_3 = -0.16$ | $a_4 = 0.05$ |
|---|---|---|---|---|
| $0.30^2$ | 0.3979 | -0.2416 | -0.1627 | 0.0455 |
| $0.80^2$ | 0.4087 | -0.2432 | -0.1752 | 0.0444 |
| $1.0^2$ | 0.4202 | -0.2284 | -0.1775 | 0.0328 |
| | $b_{11} = -0.45$ | $b_{12} = 0.32$ | $b_{13} = 0.18$ | $b_{14} = -0.10$ |
| $0.30^2$ | -0.4488 | 0.3162 | 0.1838 | -0.0832 |
| $0.80^2$ | -0.4527 | 0.3139 | 0.1716 | -0.0779 |
| $1.0^2$ | -0.4599 | 0.3094 | 0.1738 | -0.0685 |
| | $b_{21} = -0.02$ | $b_{22} = 0.10$ | $b_{23} = -0.07$ | $b_{24} = 0.00$ |
| $0.30^2$ | -0.0209 | 0.1059 | -0.0599 | -0.0283 |
| $0.80^2$ | -0.0251 | 0.1026 | -0.0497 | -0.0457 |
| $1.0^2$ | -0.0349 | 0.1024 | -0.0393 | -0.0447 |
| | $b_{31} = 0.40$ | $b_{32} = -0.05$ | $b_{33} = 0.00$ | $b_{34} = 0.20$ |
| $0.30^2$ | 0.3890 | -0.0504 | -0.0126 | 0.1987 |
| $0.80^2$ | 0.3945 | -0.0542 | -0.0032 | 0.1871 |
| $1.0^2$ | 0.3868 | -0.0440 | -0.0098 | 0.1721 |
| | $b_{41} = 0.05$ | $b_{42} = 0.00$ | $b_{43} = 0.30$ | $b_{44} = -0.20$ |
| $0.30^2$ | 0.0502 | 0.0014 | 0.2918 | -0.1876 |
| $0.80^2$ | 0.0604 | -0.0013 | 0.2903 | -0.1817 |
| $1.0^2$ | 0.0638 | 0.0050 | 0.2836 | -0.1806 |
| | $f_1 = 1.20$ | $f_2 = 1.60$ | $f_3 = 0.60$ | $f_4 = 2.12$ |
| $0.30^2$ | 1.1939 | 1.5777 | 0.5998 | 2.1146 |
| $0.80^2$ | 1.1890 | 1.5758 | 0.6074 | 2.1112 |
| $1.0^2$ | 1.1869 | 1.5818 | 0.6378 | 2.1389 |
| | $k = -0.41$ | Total parameters estimation error ($\delta_\theta$%) | | |
| $0.30^2$ | -0.4102 | 1.5298 | | |
| $0.80^2$ | -0.3987 | 2.2598 | | |
| $1.0^2$ | -0.3938 | 3.1516 | | |

**Table 4:** The parameter estimates and errors of the B-PF-RLS algorithm for the fourth-order system

## Example 3

In this context, we evaluate the B-PF-RLS algorithm using two-tank model shown in Figure 13. From a practical standpoint, the simulation aims to demonstrate the algorithm's effectiveness.

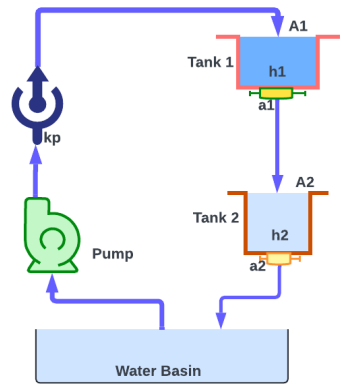

**Figure 13:** Schematic of the state coupled two-tanks system.

The tank process is a pilot tank system comprising two water tanks connected in cascade. Using physical modelling based on Torricelli's principle and the net change of volume in the tank, the system can be described by the following nonlinear model [30, 31].

$$\dot{h}_1 = -\frac{a_1\sqrt{2g}}{A_1}\sqrt{h_1} + \frac{k_p}{A_1}u \tag{20}$$

$$\dot{h}_2 = \frac{a_1\sqrt{2g}}{A_2}\sqrt{h_1} - \frac{a_2\sqrt{2g}}{A_2}\sqrt{h_2} \tag{21}$$

Where the input signal, $u$ is the voltage to the electrical pump, and $h_1$ and $h_2$ are the water levels (in cm) in the upper and lower tanks, respectively. $A_1$ and $A_2$ represent the areas of the upper and lower tanks. The effluent areas are denoted by $a_1$ and $a_2$, and $g$ and $k_p$ are the gravity constant and pump constant respectively. The nonlinear system above can be further simplified by making a linearization around a working point. Working level considered here is $L_{10} = 0.02556 cm$, $L_{12} = 0.0567 cm$ about which is derived from the linearized model. Defining the output signals as the voltages $h_1$ and $h_2$ (in volts) from the water level sensors leads to the following linear state-space model [30]. The two tank model parameters are presented in Table 5.

$$\begin{bmatrix}\dot{h}_1 \\ \dot{h}_2\end{bmatrix} = \begin{bmatrix} -\frac{a_1 g\sqrt{2}}{2A_1\sqrt{gL_{10}}} & 0 \\ \frac{a_1 g\sqrt{2}}{2A_1\sqrt{gL_{10}}} & \frac{-a_2 g\sqrt{2}}{2A_2\sqrt{gL_{20}}} \end{bmatrix}\begin{bmatrix}h_1 \\ h_2\end{bmatrix} + \begin{bmatrix}\frac{k_p}{A_1} \\ 0\end{bmatrix}u(t) \tag{22}$$

$$y(t) = \begin{bmatrix}0 & 1\end{bmatrix}\begin{bmatrix}h_1 \\ h_2\end{bmatrix} \tag{23}$$

Studying the linearized model with some modifications reveals that for the system Tank 1, the first order model will have $u$ and $h_1$ as single input and output, respectively. However, for the system Tank 2 the input is given by both $u$ and $h_1$, whereas it has a single output $h_2$ see Figure 14. Hence, a cascade structure in Figure 14 gives the following second order bilinear system with some modifications to consider $u(t)$ is the input and $h_2$ is the output [32].

$$\begin{bmatrix}\dot{x}_1 \\ \dot{x}_2\end{bmatrix} = \begin{bmatrix} \frac{-a_2 g\sqrt{2}}{2A_2\sqrt{gL_{20}}} & \frac{a_1 g\sqrt{2}}{2A_1\sqrt{gL_{10}}} \\ 0 & -\frac{a_1 g\sqrt{2}}{2A_1\sqrt{gL_{10}}} \end{bmatrix}\begin{bmatrix}x_1 \\ x_2\end{bmatrix} + \begin{bmatrix}b_{22} & b_{21} \\ b_{12} & b_{11}\end{bmatrix}\begin{bmatrix}x_1 \\ x_2\end{bmatrix}u(t) + \begin{bmatrix}0 \\ \frac{k_p}{A_1}\end{bmatrix}u(t)$$

$$y(t) = \begin{bmatrix}1 & 0\end{bmatrix}\begin{bmatrix}x_1 \\ x_2\end{bmatrix}$$

Here, $x_1 = h_2$ and $x_2 = h_1$

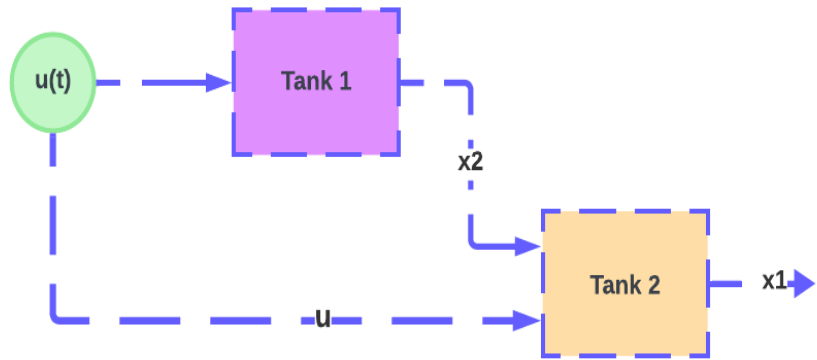

**Figure 14:** Subprocess for two-tank bilinear model

To streamline the simulation process, we employ the original model and discretize it using the forward Euler method. As a result, and after substituting the tanks parameters according to Table 5 and transform the model into its observable canonical form the bilinear model to be identified will be

$$\begin{bmatrix}\dot{x}_1\\\dot{x}_2\end{bmatrix}=\begin{bmatrix}-0.2773 & 1\\0.0190 & 0\end{bmatrix}\begin{bmatrix}x_1\\x_2\end{bmatrix}+\begin{bmatrix}-0.007 & 0.0090\\0 & 0.0095\end{bmatrix}\begin{bmatrix}x_1\\x_2\end{bmatrix}u(t)+\begin{bmatrix}0.035\\0.0092\end{bmatrix}u(t)+w(t)$$

$$y(t)=\begin{bmatrix}1 & 0\end{bmatrix}\begin{bmatrix}x_1\\x_2\end{bmatrix}+0.28(v-1)+v(t)$$

The parameter vector to be identified is
$$\theta=[a_1,a_2,b_1,b_1,b_1,b_1,f_1,f_2,J_1]^T$$
$$=[-0.2773, 0.0190, -0.007, 0.0090, 0, 0.0095, 0.035, 0.0092, 0.28]^T$$

- In the simulation, the input $\{u(t)\}$ is a pseudo-random binary sequence generated by the MATLAB function modified to have varying amplitudes. ,$w_1(t)$ and $w_2(t)$ are random noise sequences with zero mean and variance $\sigma^2_{w_1}=0.05^2$, and $\sigma^2_{w_2}=0.001^2$ respectively. $v(t)$ is a random noise sequence with zero mean and variance $\sigma^2_v=0.60^2$ , $\sigma^2_v=0.80^2$ and $\sigma^2_v=1.0^2$ . Set the data length $L=5000$ and Generate system parameter and state estimates by applying the B-PF-RLS algorithm with different number of particles according to $\sigma^2_v$ selection.
- The parameters estimates and errors $\delta_\theta=\|\hat{\theta}-\theta\|/\|\theta\|$ at $\sigma^2_v=0.60^2$ , $\sigma^2_v=0.8^2$ and $\sigma^2_v=1.0^2$ are summarized in Table 6. Specified value of the number of particles are chosen, and the parameter estimation errors are plotted against $t$ for $\sigma^2_v=0.60^2$ , $\sigma^2_v=0.80^2$ and $\sigma^2_v=1.0^2$ in Fig. (15).
- Without knowing the measurement noise variance, as discussed in section 2.2, Table 7 presents the parameter estimation and error using the proposed algorithm (55)-(65), (69)-(74) with unknown measurement noise. The process noise variances are $\sigma^2_{w_1}=0.05^2$, and $\sigma^2_{w_2}=0.001^2$.
- To further validate the effectiveness of the model obtained through the B-PF-RLS algorithm, a different data set consisting of 100 samples ($L_r=100$) from $t=L+1$ to $t=L+L_r$ was used. The parameter estimates from the fifth row in Table 6 were employed to construct the resulting model. Figure 14 presents the actual output $y(t)$, the predicted output $\hat{y}(t)$, and their errors $\hat{y}(t)-y(t)$. The figure demonstrates that the predicted output closely tracks the actual output with high accuracy and minimal errors.

| $A_1$ | $A_2$ | $a_1$ | $a_2$ | $k_p$ | $L_{10}$ | $L_{20}$ |
|---|---|---|---|---|---|---|
| $3.80\ cm^2$ | $2.4697 cm^2$ | $0.57\ cm^2$ | $0.50\ cm^2$ | $5.88\ cm^3/volts-sec$ | $0.02556\ cm$ | $0.05676\ cm$ |

**Table 5:** Tanks Parameter values

| $\sigma_v^2$ | t | $a_1$ | $a_2$ | $b_{11}$ | $b_{12}$ | $b_{21}$ | $b_{22}$ | $f_1$ | $f_2$ | k | $\delta_\theta$% |
|---|---|---|---|---|---|---|---|---|---|---|---|
| $0.60^2$ | 100 | -0.0903 | 0.0502 | -0.0007 | 0.0005 | -0.0034 | 0.0034 | 0.0200 | 0.0389 | 0.0750 | 70.9952 |
|  | 1000 | -0.2517 | 0.0647 | -0.0048 | 0.0015 | -0.0044 | 0.0083 | 0.0362 | 0.0126 | 0.2670 | 13.8302 |
|  | 2000 | -0.2548 | 0.0537 | -0.0059 | 0.0016 | -0.0036 | 0.0087 | 0.0337 | 0.0128 | 0.2823 | 10.6990 |
|  | 5000 | -0.2777 | 0.0182 | -0.0047 | 0.0017 | -0.0009 | 0.0090 | 0.0359 | 0.0085 | 0.2878 | 2.7960 |
| $0.80^2$ | 100 | -0.0978 | 0.0135 | -0.0038 | -0.0009 | -0.0022 | 0.0030 | 0.0227 | 0.0324 | 0.1025 | 64.1118 |
|  | 1000 | -0.2269 | 0.0107 | -0.0034 | -0.0010 | -0.0025 | 0.0026 | 0.0342 | 0.0139 | 0.2304 | 18.2894 |
|  | 2000 | -0.2518 | 0.0209 | -0.0041 | -0.0015 | -0.0028 | 0.0030 | 0.0360 | 0.0093 | 0.2570 | 9.2743 |
|  | 5000 | -0.2729 | 0.0085 | -0.0041 | -0.0010 | -0.0012 | 0.0032 | 0.0365 | 0.0078 | 0.2744 | 4.4821 |
| $1.0^2$ | 100 | -0.0537 | 0.0170 | 0.0021 | 0.0005 | -0.0015 | -0.0009 | 0.0138 | 0.0434 | 0.0382 | 83.7903 |
|  | 1000 | -0.2245 | 0.0642 | -0.0015 | 0.0006 | -0.0037 | -0.0004 | 0.0360 | 0.0138 | 0.2534 | 19.1713 |
|  | 2000 | -0.2278 | 0.0621 | -0.0030 | 0.0006 | -0.0034 | -0.0003 | 0.0334 | 0.0147 | 0.2756 | 17.0136 |
|  | 5000 | -0.2514 | 0.0211 | -0.0021 | 0.0005 | -0.0003 | -0.0003 | 0.0362 | 0.0098 | 0.2861 | 7.5785 |
| True value |  | -0.2773 | 0.0190 | -0.0070 | 0.0090 | 0.0000 | 0.0095 | 0.0350 | 0.0092 | 0.2800 |  |

**Table 6** The parameter estimates and errors of the B-PF-RLS algorithm for $\sigma_v^2 = 0.60^2$, $\sigma_v^2 = 0.80^2$ and $\sigma_v^2 = 1.0^2$

| $\sigma_v^2$ | t | $a_1$ | $a_2$ | $b_{11}$ | $b_{12}$ | $b_{21}$ | $b_{22}$ | $f_1$ | $f_2$ | k | $\delta_\theta$% |
|---|---|---|---|---|---|---|---|---|---|---|---|
| $0.60^2$ (But actually unknown) | 100 | -0.0803 | 0.0423 | -0.0009 | 0.0005 | -0.0027 | 0.0033 | 0.0198 | 0.0399 | 0.0691 | 73.6147 |
|  | 1000 | -0.2489 | 0.0555 | -0.0052 | 0.0014 | -0.0038 | 0.0080 | 0.0364 | 0.0124 | 0.2726 | 12.0496 |
|  | 2000 | -0.2509 | 0.0479 | -0.0061 | 0.0014 | -0.0032 | 0.0082 | 0.0339 | 0.0127 | 0.2858 | 10.2467 |
|  | 5000 | -0.2738 | 0.0160 | -0.0047 | 0.0014 | -0.0008 | 0.0082 | 0.0361 | 0.0085 | 0.2901 | 3.4725 |
| True value |  | -0.2773 | 0.0190 | -0.0070 | 0.0090 | 0.0000 | 0.0095 | 0.0350 | 0.0092 | 0.2800 |  |

**Table 7** The parameter estimates and errors of the B-PF-RLS algorithm for $\sigma_v^2$ actually unknown

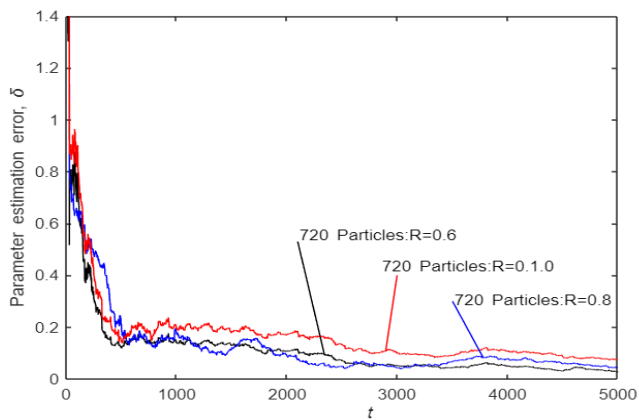

**Figure 15:** The B-PF-RLS estimation errors – against t with $\sigma_v^2 = 0.60^2$, $\sigma_v^2 = 0.80^2$ and $\sigma_v^2 = 1.0^2$

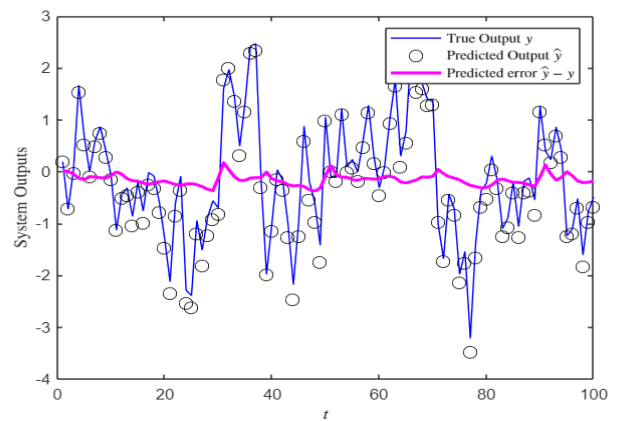

**Figure 16:** The true outputs and the predicted outputs for the bilinear two tank system

## 6. Conclusion

In this study, we proposed a comprehensive algorithm for jointly estimating the parameters and states of a bilinear system affected by colored measurement noise. The proposed approach integrates a Recursive Least Squares (RLS) estimator for parameter identification and a Particle Filter (PF) for state estimation, effectively addressing the challenges posed by the system's nonlinear characteristics and the presence of noise. Through extensive simulations, including a two-tank system model and a higher-order system, we demonstrated the robustness and accuracy of our B-PF-RLS algorithm. The results confirm that our approach can reliably estimate both parameters and states even under challenging conditions, such as when process noise is unknown. This demonstrates the versatility and effectiveness of the algorithm in a variety of practical scenarios. Future work may explore further refinements to the algorithm to enhance its performance in real-time applications and investigate its applicability to other nonlinear systems. Additionally, the integration of more sophisticated noise modelling techniques could further improve estimation accuracy.

## References

1. Otto, S., Peitz, S., & Rowley, C. (2024). Learning Bilinear Models of Actuated Koopman Generators from Partially Observed Trajectories. *SIAM Journal on Applied Dynamical Systems, 23*(1), 885-923.
2. Liu, S. (2017). An overview of chemical reaction analysis.
3. Hu, T. (2010). A nonlinear-system approach to analysis and design of power-electronic converters with saturation and bilinear terms. *IEEE transactions on power electronics, 26*(2), 399-410.
4. Benesty, J., Paleologu, C., Dogariu, L. M., & Ciochină, S. (2021). Identification of linear and bilinear systems: A unified study. *Electronics, 10*(15), 1790.
5. Li, M., & Liu, X. (2018). The least squares based iterative algorithms for parameter estimation of a bilinear system with autoregressive noise using the data filtering technique. *Signal Processing, 147*, 23-34.
6. Gu, Y., Zhu, Q., Liu, J., Zhu, P., & Chou, Y. (2020). Multi-Innovation Stochastic Gradient Parameter and State Estimation Algorithm for Dual-Rate State-Space Systems with d-Step Time Delay. *Complexity, 2020*(1), 6128697.
7. Zhang, X., & Ding, F. (2020). Hierarchical parameter and state estimation for bilinear systems. *International Journal of Systems Science, 51*(2), 275-290.
8. Li, M., & Liu, X. (2018). Auxiliary model based least squares iterative algorithms for parameter estimation of bilinear systems using interval-varying measurements. *IEEE access, 6*, 21518-21529.
9. Li, M., Liu, X., & Ding, F. (2017). The maximum likelihood least squares based iterative estimation algorithm for bilinear systems with autoregressive moving average noise. *Journal of the Franklin Institute, 354*(12), 4861-4881.
10. Liu, S., Ding, F., & Hayat, T. (2020). Moving data window gradient-based iterative algorithm of combined parameter and state estimation for bilinear systems. *International Journal of Robust and Nonlinear Control, 30*(6), 2413-2429.
11. Zhang, X., Ding, F., Xu, L., Alsaedi, A., & Hayat, T. (2019). A hierarchical approach for joint parameter and state estimation of a bilinear system with autoregressive noise. *Mathematics, 7*(4), 356.
12. Wang, X., Ma, J., & Xiong, W. (2023). Expectation-maximization Estimation Algorithm for Bilinear State-space Systems with Missing Outputs Using Kalman Smoother. *International Journal of Control, Automation and Systems, 21*(3), 912-923.
13. Phan, M. Q., Vicario, F., Longman, R. W., & Betti, R. (2015). Optimal bilinear observers for bilinear state-space models by interaction matrices. *International Journal of Control, 88*(8), 1504-1522.
14. Sacchelli, L., Brivadis, L., Serres, U., & Yaacov, I. B. (2024). Output feedback stabilisation of bilinear systems via control templates. *arXiv preprint arXiv:2403.01869.*
15. Hara, S., & Furuta, K. (1976). Minimal order state observers for bilinear systems. International Journal of Control, 24(5), 705-718.
16. Derese, I., Stevens, P., & Noldus, E. (1979). Observers for bilinear systems with bounded input. *International Journal of Systems Science, 10*(6), 649-668.
17. Ying, Y. Q., Sun, Y. X., & Rao, M. (1991). Bilinear state-disturbance composite observer and its application. *International journal of systems science, 22*(12), 2489-2498.
18. Zhang, X., Liu, Q., Ding, F., Alsaedi, A., & Hayat, T. (2020). Recursive identification of bilinear time-delay systems through the redundant rule. *Journal of the Franklin Institute, 357*(1), 726-747.
19. Li, M., Liu, X., & Ding, F. (2017). The gradient-based iterative estimation algorithms for bilinear systems with autoregressive noise. *Circuits, Systems, and Signal Processing, 36*, 4541-4568.
20. Poterjoy, J. (2016). A localized particle filter for high-dimensional nonlinear systems. *Monthly Weather Review, 144*(1), 59-76.
21. Zhang, X., Xu, L., Ding, F., & Hayat, T. (2018). Combined state and parameter estimation for a bilinear state space system with moving average noise. *Journal of the Franklin Institute, 355*(6), 3079-3103.
22. Liu, S., Zhang, Y., Xu, L., Ding, F., Alsaedi, A., & Hayat, T. (2021). Extended gradient-based iterative algorithm for bilinear state-space systems with moving average noises by using the filtering technique. *International Journal of Control, Automation and Systems, 19*(4), 1597-1606.
23. Zhang, X., Ding, F., Xu, L., & Yang, E. (2018). State filtering-based least squares parameter estimation for bilinear systems using the hierarchical identification principle. *IET Control Theory & Applications, 12*(12), 1704-1713.
24. Zhang, J., & Lee, J. (2011). A review on prognostics and health monitoring of Li-ion battery. *Journal of power sources, 196*(15), 6007-6014.
25. Chen, J., Liu, Y., Ding, F., & Zhu, Q. (2018). Gradient-based particle filter algorithm for an ARX model with nonlinear

26. Liu, S., Zhang, Y., Xu, L., Ding, F., Alsaedi, A., & Hayat, T. (2021). Extended gradient-based iterative algorithm for bilinear state-space systems with moving average noises by using the filtering technique. *International Journal of Control, Automation and Systems, 19*(4), 1597-1606.
27. Liu, S., Ding, F., & Yang, E. (2021). Iterative state and parameter estimation algorithms for bilinear state-space systems by using the block matrix inversion and the hierarchical principle. *Nonlinear Dynamics, 106*(3), 2183-2202.
28. Hag ElAmin, K. A. E. M. (2020). Clustering Input Signals Based Identification Algorithms for Two-Input Single-Output Models with Autoregressive Moving Average Noises. *Complexity, 2020*(1), 2498487.
29. A. El Mageed Hag Elamin Khalid, "Perspective Chapter: Insights from Kalman Filtering with Correlated Noises Recursive Least-Square Algorithm for State and Parameter Estimation," in Kalman *Filters -Theory, Applications, and Optimization [Working Title],* IntechOpen, 2024.
30. G Stohy, M., S Abbas, H., M El-Sayed, A. H., & G Abo El-maged, A. (2020). Parameter estimation and PI control for a water coupled tank system. *Journal of Advanced Engineering Trends, 38*(2), 147-159.
31. Folorunso, T. A., Bello-Salau, H., Olaniyi, O. M., & Abdulwahab, N. (2013). Control of a two layered coupled tank: Application of IMC, IMC-PI and pole-placement PI controllers.
32. Ekman, M. (2005). *Modeling and control of bilinear systems: application to the activated sludge process* (Doctoral dissertation, Acta Universitatis Upsaliensis).